\documentstyle[preprint,aps,epsf]{revtex}
\begin{document}

\title{ Multifragmentation and the Phase Transition:
A Systematic Study of the MF of  1A GeV Au, La, and Kr
}
\author{B. K. Srivastava$^1$,  R. P. Scharenberg$^1$,
S. Albergo$^2$, F. Bieser$^6$, F. P. Brady$^3$, 
Z. Caccia$^2$, D. A. Cebra$^3$, A. D. Chacon$^7$, 
J. L. Chance$^3$, Y. Choi$^ 1$,
S. Costa$^2$, J. B. Elliott$^1$, M. L. Gilkes$^1$, J. A. Hauger$^1$,
A. S. Hirsch$^1$, E. L. Hjort$^1$,
A. Insolia$^2$, M. Justice$^5$, D. Keane$^5$, 
J. C. Kintner$^3$, V. Lindenstruth$^4$,
M. A. Lisa$^6$, H. S. Matis$^6$, M. McMahan$^6$, C. McParland$^6$,
W. F. J. M\"{u}ller$^4$,
D. L. Olson$^6$, M. D. Partlan$^3$, N. T. Porile$^1$, R. Potenza$^2$,
G. Rai$^6$, J. Rasmussen$^6$, H. G. Ritter$^6$, J. Romanski$^2$, 
J. L. Romero$^3$, G. V. Russo$^2$,
H. Sann$^4$, A. Scott$^5$, 
Y. Shao$^5$, T. J. M. Symons$^6$, M. Tincknell$^1$,
C. Tuv\'{e}$^2$, S. Wang$^5$, P. Warren$^1$, H. H. Wieman$^6$,
T. Wienold$^6$, and K. Wolf$^7$\\
(EOS Collaboration)}
\address{$^1$Purdue University, West Lafayette, IN 47907 \\
$^2$Universit\'{a} di Catania and Istituto Nazionale di Fisica
Nucleare-Sezione di Catania,\\
95129 Catania, Italy \\
$^3$University of California, Davis, CA 95616\\
$^4$GSI, D-64220 Darmstadt, Germany \\
$^5$Kent State University, Kent, OH 44242 \\
$^6$Nuclear Science Division, Lawrence Berkeley National Laboratory, 
Berkeley, CA 94720 \\
$^7$Texas A\&M University, College Station, TX  77843}
\date{\today}

\maketitle
\begin{abstract}
A systematic analysis of the multifragmentation (MF)  in fully reconstructed events 
from 1A GeV Au, La and Kr collisions with C has been performed. 
This data is used to provide a definitive test of the variable
 volume version of the statistical multifragmentation model (SMM).  
A single set of SMM parameters directly determined by the data and the semi-empirical
mass formula are used after the adjustable inverse level density parameter,
$\epsilon_{o}$ is determined 
by the fragment distributions. The results from SMM for second stage multiplicity, 
size of the biggest fragment and the intermediate mass fragments are in excellent agreement with the data. 
 Multifragmentation thresholds have been obtained for all three systems using SMM prior to secondary decay. The data indicate that both thermal excitation energy   
$E_{th}^{*}$ and the isotope ratio temperature $T_{He-DT}$ 
decrease with increase in system size at the critical point. The breakup temperature obtained from SMM also shows the same trend as seen in the data. 
The SMM model is used to study the nature of the MF phase transition. The caloric curve 
for Kr exhibits back-bending ( finite latent heat) while the caloric curves for Au and La 
are consistent with a continuous phase transition ( nearly zero latent heat)
and the values of 
the critical exponents $\tau$, $\beta$ and $\gamma$, both from
data and SMM, are close to those for a  'liquid-gas' 
system for Au and La.
We conclude that the larger Coulomb expansion energy in Au and La reduces the latent heat 
required for MF and changes the nature of the phase transition. Thus the Coulomb energy
plays a major role in nuclear MF.

\end{abstract}
\vspace{24pt}
\newpage
\section{INTRODUCTION}
 
 During the past decade a large effort has been made to understand  the
multifragmentation (MF) process in heavy ion reactions. Recently a number of
review articles have appeared describing the details of this phenomenon\cite{gross1,moretto93,bondorf3,gross2,richert00,bonasera00,tsang00}. 
 The Purdue group was the first to suggest that nuclear MF might be
a critical phenomenon - a second order phase transition occurring near a critical point\cite{finn82,minich82}. 
 In an inclusive experiment performed at Fermilab, the spectra of mass identified fragments resulting from proton interactions 
with both Kr and Xe gas targets were measured \cite{finn82,minich82}. 
The discovery that the yields of fragments with mass $\it A_{f}$ produced in 
p+Kr and p+Xe reactions, with proton beam energy from 80 to 350 GeV,
obeyed a power law $ Y(A_{f})\sim  A_{f}^{-\tau}$, with $\tau \sim 2.5$
\cite{minich82} generated theoretical interest in MF
 in terms of a continuous phase transition. A similar power 
law was also predicted by Fisher \cite{fisher67} for a mass distribution
of droplets at a liquid-gas phase transition critical point. Thus the 
results from the Purdue work gave a hint that MF could provide important
information about the equation of state of nuclear matter \cite{siemens83,schlagel87,muller95,chase95}. The analysis of the fragment kinetic energy 
spectra suggested that the fragments are emitted 
from a less than normal density system
  in the decay of a common remnant which is lighter 
than the target\cite{minich82,poskanzer71,westfall78,hirsch84}.
An analysis of the fragment yields based on a thermal liquid drop
model gave a freeze-out temperature of $\sim $ 5 MeV 
\cite{minich82,hirsch84}. 
Similar results were observed when the above study was extended to lower energies\cite{mahi88,porile89}. This view was supported by 
exclusive emulsion MF data \cite{waddington85}, which were 
analyzed by Campi to show that the conditional moments of the individual 
fragment events exhibited characteristics of a phase transition 
\cite{campi1,campi2,campi3}. Bauer and Campi were the first to apply the
 methods used in percolation
studies to analyze MF data \cite{campi1,campi2,campi3,bauer1,bauer2}. In percolation theory the moments of the cluster distribution contain the signature of critical behavior \cite{stauffer79,stauffer92}.

 In recent years further progress was made by experiments
 in which practically all
the fragments emitted in a given event were detected,
thereby permitting complete reconstruction of MF events \cite{hubele91,kreutz93,schuttauf96,orion,isis,isis0,isis1,isis2}.
 The ALADIN Collaboration 
studied the MF of 400A-1000A MeV Xe, Au, and U nuclei on various targets. Their
results showed that fragment yields were independent of the entrance channel
when the data were scaled for projectile or target mass \cite{hubele91,kreutz93,schuttauf96}. The EOS Collaboration 
studied the MF of 1A GeV gold on carbon
and analyzed the data using methods developed in the study of critical 
phenomena
\cite{gilkes94,hauger96,elliott96,hauger98,elliott98,lauret98,sriv99}.
 Several critical exponents were determined and their values suggested that 
the MF of Au can be 
understood as due to a continuous phase transition \cite{elliott5,elliott6}.  
The first results of the MF of 1A GeV La and Kr on carbon have also been
reported \cite{hauger00,srivas00}. 
The MF transition appears to
involve the breakup of a nucleus to form several IMF's. The production
of nucleons and light particles - the nuclear analog of a gas- occurs largely 
in the fragment deexcitation step\cite{scharn2}.
Along with the earlier inclusive studies, these experiments 
suggested that prior to MF the remnant formed in the prompt stage
achieves thermal equilibrium. 
A two step process was proposed for the collision. This two step 
 process is an idealization of a time dependent process. In the first 
prompt stage nucleons are knocked out of the participants. The emission 
of these prompt particle leaves an equilibrated remnant nucleus 
which undergoes deexcitation in a second step.

In the EOS experiment prompt stage was separated from the MF stage by making a cut on the kinetic energy of light charged particles\cite{hauger98}. This
separation was made possible by the ability of the EOS detector to
measure nearly all the charged particles and fragments emitted in each event 
using reverse kinematics\cite{hauger98}. Complete reconstruction and stage separation was 
therefore possible for a majority of the events. These reconstructed events 
were characterized by mass, charge, and excitation energy of the remnant. 
The resultant  remnant then undergoes MF to produce excited fragments which then undergo 
secondary decay to produce the observed fragments \cite{scharn2}.
The ALADIN experiment is also capable of measuring all the charged particles except for Z=1 particles\cite{hubele91,kreutz93,schuttauf96}. The ISiS
Collaboration has also produced one of the most complete MF data sets in the high energy collision of $\it p$, $\it \bar p$, and $\pi^-$ with gold \cite{isis,isis0,isis1,isis2}. In the ISiS experiment the charge and mass of the excited source were obtained on an event by event basis by subtracting the non equilibrium particles from the target charge and mass. However, the  non-equilibrium 
particles could only be separated from the thermal particles by means of a 
 parametrization involving two component moving source fits.

 The  EOS\cite{hauger98} data can be compared with statistical
models. There are several statistical models which 
have been used to study MF, 
\cite{gross1,bondorf3,gross2,bondorf1,bondorf2,botvina87,friedman90,durand92,chase94,chase97,lee97,dasgupta98} but the most 
widely used are the Statistical Multifragmentation Model (SMM)
\cite{bondorf3,bondorf1,bondorf2,botvina87}
and Monte Carlo microcanonical model (MMMC) \cite{gross1,gross2}.
The theoretical interest in MF is not confined to only statistical models but 
several other approaches to study MF have been carried out e.g. percolation,
lattice gauge and Ising models \cite{campi2,bauer88,pan95,campi97,dasgupta97,pan98,carmona98,gulm99}. A phenomenological droplet model \cite{goodman84},
based on the Fisher\cite{fisher67} droplet model, has also been used to describe the liquid-gas phase transition in nuclear reactions.  
Here we shall compare the
experimental data with the predictions of the SMM model 
in the manner presented in previous publications \cite{hauger00,scharn2,scharn1,sriv1}. We found that this model is in good agreement with a
variety of results for the MF of 1A GeV Au \cite{scharn2}. SMM requires
mass, charge and excitation energy of the remnant as input.
Using the experimental remnants from Au+C, the parameters of the SMM 
 were fixed based on the 
agreement between SMM and data. 
Some comparisons with the data for La+C and Kr+C were also made in later work using the same set of parameters as in case of Au+C \cite{hauger00}.

 This paper deals with the MF of 1A GeV Au, La, and Kr on carbon. Section II gives a brief summary of the experimental remnant properties. The fragment properties 
are discussed in Section III. Section IV describes SMM and gives a 
comparison with the data. The search for the critical transition is discussed in Section V. The determination of various critical exponents  is given in Section VI.
 Energy fluctuations and heat capacity analyses are discussed in Section VII.
Section VIII dwells on the nature of the phase transition in Au, La and Kr. Conclusions 
are given in Section IX.

\section{ PROPERTIES OF THE REMNANT}
 The reverse kinematic EOS experiment was performed with 1A GeV 
$^{197}$Au, $^{139}$La, and $^{84}$Kr
beams on carbon targets. The details of the experiment are given in our earlier publications \cite{hauger98,hauger00}. Only a brief description will be given 
here.  
 The experiment was done with the EOS Time Projection Chamber (TPC)\cite{rai90,wieman91} and a multiple sampling ionization chamber (MUSIC II)\cite{bauer97}.
The TPC provided almost 4 $\pi$ solid angle coverage in the center-of-mass system. Three-dimensional tracking and charged particle identification permitted
momentum and energy reconstruction of fragments with charges in the range of 
$1\leq Z \leq 8$. Particle identification was based on specific energy loss
along particle tracks. MUSIC II detected and tracked fragments with 
charges $8 \leq Z \leq Z_{beam}$. The excellent charge resolution of this 
detector permitted identification of all detected fragments. 

 The analysis presented here is based on fully reconstructed MF events for 
which the total charge of the system was taken as $79 \leq Z \leq 83$,
 $54 \leq Z \leq 60$, and $33 \leq Z \leq 39$ for Au, La, and Kr, 
respectively \cite{hauger98,hauger00}. Approximately $\sim$ 32000, 26000, and
42000 events met the above criteria  for Au, La and Kr, respectively.

\begin{figure}[ht]
\epsfxsize=8.5cm
\centerline{\epsfbox{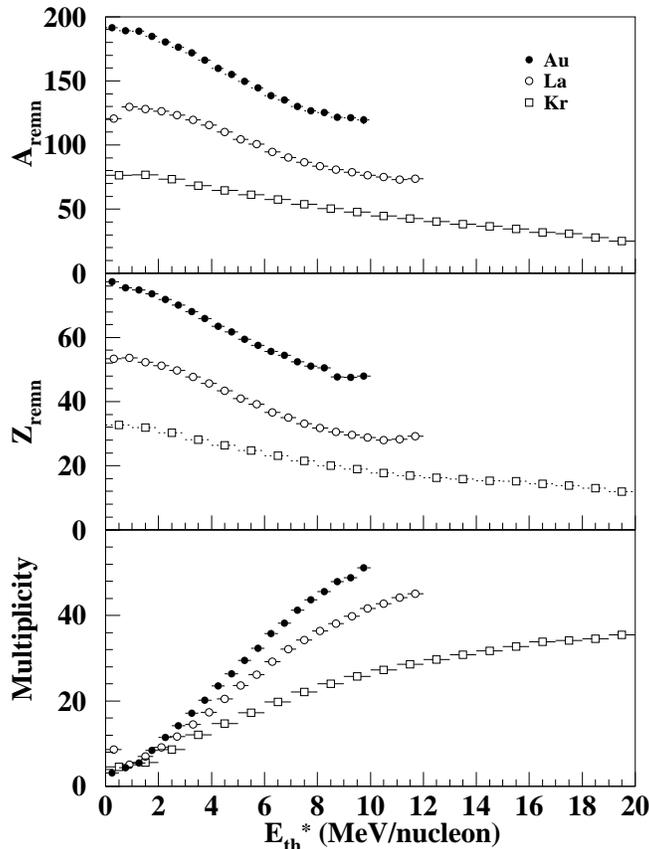}}
\caption{Average remnant mass, charge, and  total charged-particle multiplicity 
 as a function of thermal excitation energy $E_{th}^*$ from the MF of Au, La and Kr.
}
\label{auremn}
\end{figure}  
 The remnant refers to the equilibrated nucleus formed after the 
emission of prompt particles. This remnant then undergoes MF. The 
charge and mass of the remnant were obtained by removing for each event the 
total charge of the prompt particles. In order to obtain the mass, the
number of prompt neutrons was estimated by means of the ISABEL cascade 
calculation\cite{yariv1,yariv2}.  
 The excitation energy per nucleon of the remnant, $E^*$, was based on an
energy balance between the excited remnant and the final stage of the fragments\cite{cussol93} for each event, as discussed in detail elsewhere \cite{hauger98,hauger00}.

In our previous work we have shown that some of the excitation energy of the
remnant includes a  nonthermal component which may be ascribable to expansion energy, $E_x$\cite{hauger98,lauret98,hauger00}. 
The thermal excitation energy $E_{th}^*$  of the remnants is obtained as 
the difference  between $E^*$ and $E_{x}$. This energy is an important quantity
both for input to SMM and for the physics analysis of the data. Fig.\ref{auremn}  
show plots of $A_{remn}$, $Z_{remn}$ and total charged particle multiplicity, $\it m$,  
 for Au, La, and Kr, 
respectively, as a function of $E_{th}^*$ \cite{hauger98,hauger00}. 

\begin{figure}[ht]
\epsfxsize=8.5cm
\centerline{\epsfbox{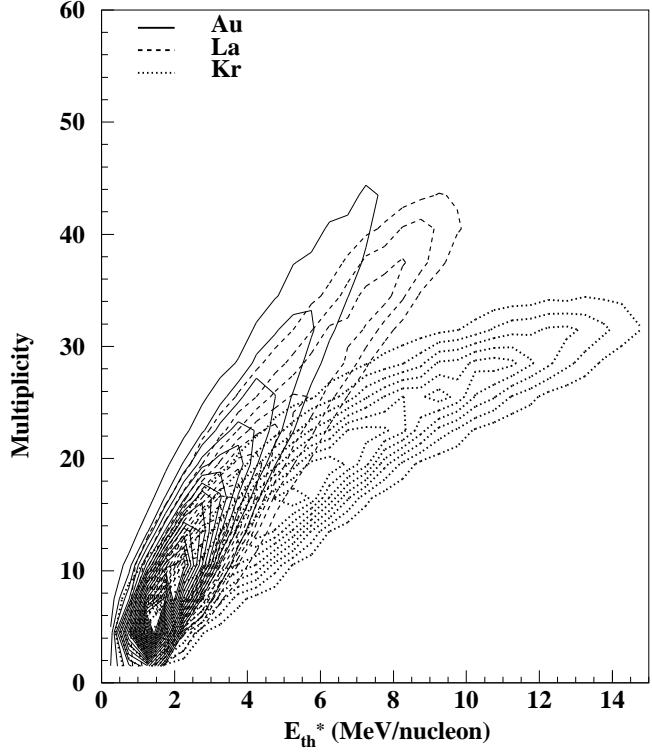}}
\caption{$\it m$ as a function of  $E_{th}^*$ for Au, La, and, Kr. There 
are ten equidistant contours.
}
\label{2dmult}
\end{figure}

The plots in Fig.\ref{auremn} show 
the nature of the remnant originating from the different system sizes. 
In case of Au and La one finds that the remnant size at the 
highest multiplicity is nearly $50\%$
 of the initial remnant size, while for Kr the remnant loses up to $\sim$ 70 $\%$ of
the initial mass. The range of $ E^{*}_{th} $ 
is also different in the three cases. 
For Kr the remnant reaches $E_{th}^*$ as high as $\sim$ 20 MeV/nucleon, 
while for Au and La  the maximum $E_{th}^*$ are $\sim 9$ and $\sim$ 
12 MeV/nucleon, respectively. Figure \ref{auremn} shows the variation of the average $\it m$
with $E_{th}^*$, but  does  not give an indication of event-to-event fluctuations. Fig.\ref{2dmult} 
shows a contour plot of $\it m$ vs $ E^{*}_{th} $. 
These two quantities are fairly closely correlated 
for Au and La and confirm the linear variation shown in Fig.\ref{auremn}. The correlation is much poorer for Kr
and a broad range of energies corresponds to a narrow range of $\it m$ values
 above 8 MeV/nucleon.

\section{FRAGMENT PROPERTIES}
\begin{figure}[ht]
\epsfxsize=8.5cm
\centerline{\epsfbox{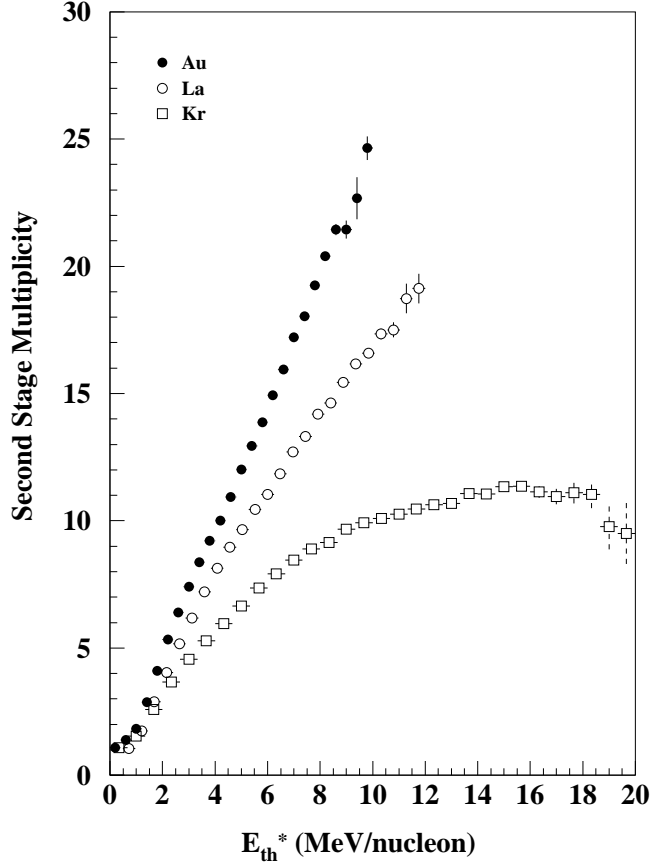}}
\caption{$\it m_{2}$ as a function of  $E_{th}^*$ for Au, La, and, Kr.
}
\label{m2eth}
\end{figure}

\begin{figure}[ht]
\epsfxsize=8.5cm
\centerline{\epsfbox{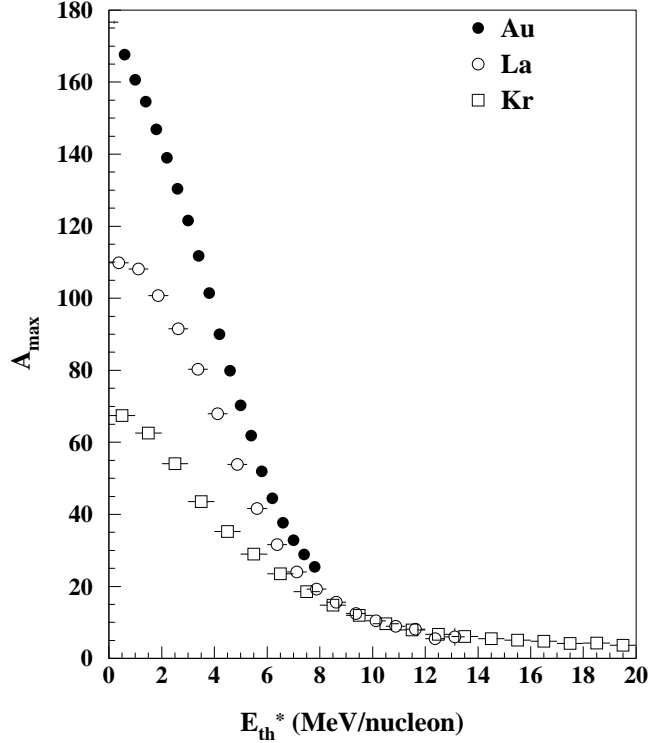}}
\caption{$A_{max}$ as a function of  $E_{th}^*$ for Au, La, and, Kr.
}
\label{amaxeth}
\end{figure}

 We first examine the  second stage fragment multiplicities,
 $\it m_{2}$. Fig.\ref{m2eth} shows a plot of  $\it m_{2}$  vs  $E_{th}^*$ for Au, La, and, Kr.  
The $\it m_{2}$ for Au and La increase linearly with $E_{th}^*$, while 
for Kr $\it m_{2}$ increases 
slowly up to $E_{th}^*$$\sim$ 8 MeV/nucleon and remains essentially constant 
at higher $E_{th}^*$. 
The constancy of  $\it m_{2}$  above 8 MeV/nucleon suggests that the 
system  may be disintegrating into individual nucleons and light particles,
which suggests that the vaporization process becomes dominant. This interpretation is confirmed 
by the variation with $E_{th}^*$ 
of the size of the largest fragment, $A_{max}$, in each event as shown in 
Fig.\ref{amaxeth}.
It is observed for Kr that above  $\sim$ 8-9 MeV/nucleon the value of 
$A_{max}$ is nearly constant with a value of $\le 6$. 
 The distribution of IMFs is shown in Fig.\ref{imfeth} for all  three systems as a function of energy. In order to exclude fission 
and/or the largest fragment for all three projctiles we define IMFs as having
nuclear charges ranging from Z=3 to Z=$Z_{proj}/4$. 
 Fig.\ref{imfeth} shows that for all the three systems the peak in IMFs is $\sim 8$ MeV/nucleon although the  IMF peak for Au is not well defined as  
$E_{th}^*$ $>$ 8 MeV/nucleon was hardly achieved for this projectile.
The drop in IMF yields for Au has been  seen when the data were plotted as a function of $\it m$ rather than $E_{th}^*$ because of the dispersion 
in $\it m$ values at
a given $E_{th}^*$ \cite{hauger00}.  Note that the number of IMFs from 
Kr drops to substantially less than 1 above 10 MeV/nucleon. This is another confirmation of the results given in the preceding section showing the 
vaporization of the Kr remnants at high $E_{th}^*$.
 IMF production has been studied as a function of projectile bombarding
energy by the ALADIN Collaboration \cite{hubele91,kreutz93,schuttauf96}. They
showed that the excitation energy dependence of the average IMF number for Xe, Au, 
and U projectiles when scaled by charge of the emitting source is similar, suggesting that
the IMF production mechanism is independent of the entrance channel. A 
universal scaling for IMF production has also been seen from a wide range of 
source masses (35 -190 nucleons) produced in reactions with energies from 
35 to 600 MeV/nucleon \cite{beaulieu96}. 

\section{THE SMM MODEL AND COMPARISON WITH DATA}

The variable volume version of SMM has been used in this work 
for comparison with the data. This variable volume corresponds approximately
to the condition of constant ( close to zero) pressure \cite{mish98}.
 SMM is a statistical description of the simultaneous breakup of an
expanded excited nucleus into nucleons and hot fragments \cite{bondorf3,bondorf1,bondorf2}. 
Individual fragments at normal nuclear density are described with a charged
liquid drop parameterization.  The free energy of a fragment 
is used to determine the fragment formation probability. This solution explicitly  assumes the inhomogeneous nature of the hot MF final state. 
\begin{figure}[ht]
\epsfxsize=8.5cm
\centerline{\epsfbox{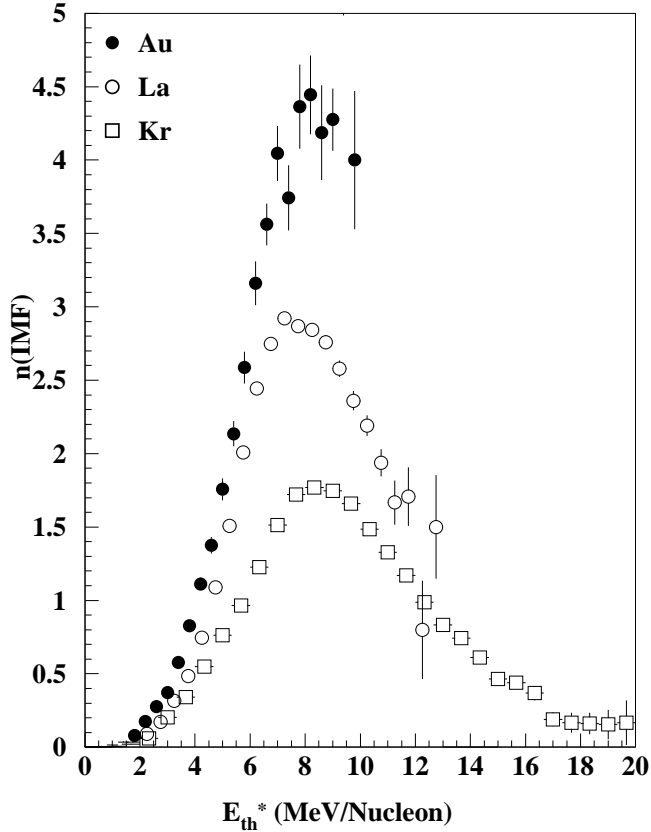}}
\caption{Average n(IMF) as a function of  $E_{th}^*$ for Au, La, and, Kr.
}
\label{imfeth}
\end{figure}

Light fragments with $Z < 3$ may also be present in the hot MF final state.
For the $Z \geq 3$ fragments, a
quantum mechanical description is used  for the temperature dependent
 volume, surface,  and translational  free energy of the fragments.
The temperature independent  parameters are based on the coefficients of the semiempirical mass formula.
The critical temperature, at which the surface tension of 
neutral nuclear matter droplets would go to zero,
is in the range suggested by infinite neutral nuclear matter calculations
\cite{ravenhall83}.  

\begin{figure}[ht]
\epsfxsize=8.5cm
\centerline{\epsfbox{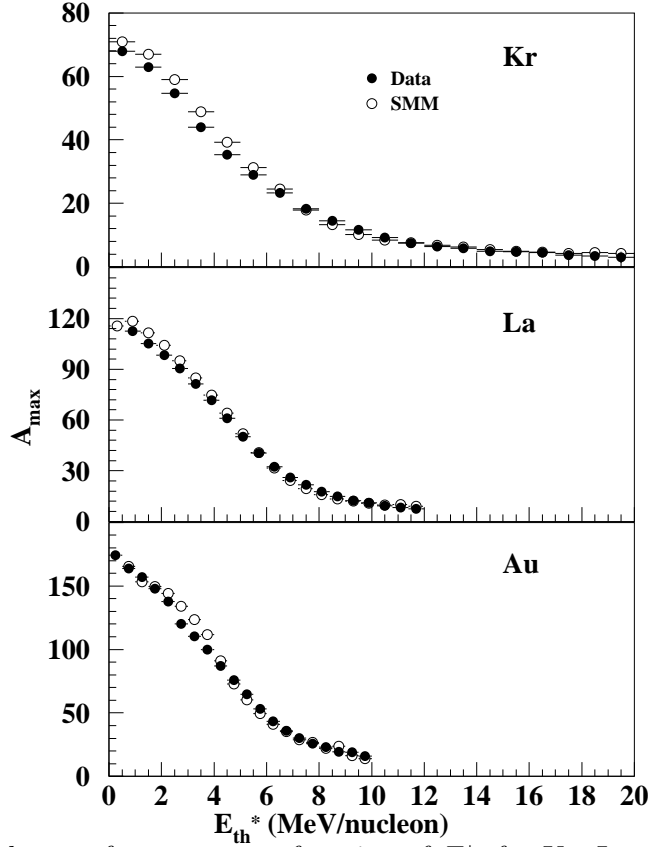}}
\caption{Size of the largest fragment as a function of  $E_{th}^*$ 
for Kr, La, and, Au from data and SMM.
}
\label{amaxethsmm}
\end{figure}

 The two important parameters of the model are the
Coulomb reduction parameter, $\kappa$,  and the inverse level density parameter
$\epsilon_{0}$ \cite{bondorf1}.
The $\kappa$ parameter was fixed with the comparison of the measured free
volume from SMM to that of the initial volume obtained in the collision 
process for the Au+C data. The details are given in ~ref.\cite{scharn2}. The
only remaining parameter, $\epsilon_{0}$, was obtained   
from the detailed  comparison of SMM results with the various 
experimental fragment properties, e.g. second stage multiplicity, size of the
biggest fragment, and IMFs using Au+C data \cite{scharn2}.
In this work the same value of $\kappa$=2 and $\epsilon_{0}$=16 MeV has been used in SMM to compare results with data for La and Kr.  
The best agreement with the data
was found by using the standard values of the parameters of the model.
\begin{figure}[ht]
\epsfxsize=8.5cm
\centerline{\epsfbox{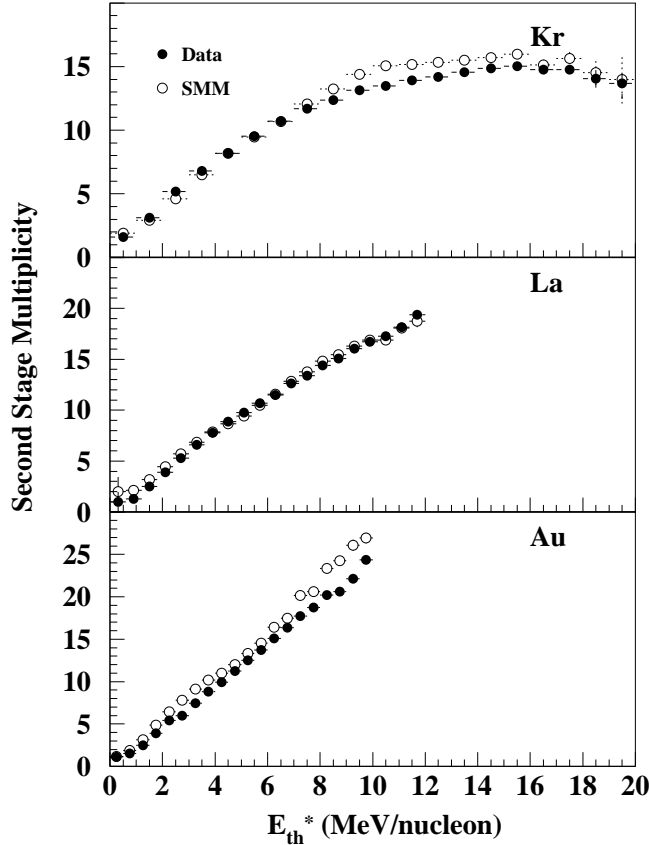}}
\caption{Second stage charged-particle multiplicity  as a function of  $E_{th}^*$ 
for Kr, La, and, Au from data and SMM.
}
\label{m2ethsmm}
\end{figure}
The SMM results were also compared with the data on La and Kr \cite{hauger00}, 
with the same parameters as in Au, using 
reduced multiplicity ($\it m/Z_{proj}$). Here we present a few comparisons
between data and SMM using $E_{th}^*$.
 Fig.\ref{m2ethsmm} shows a plot of $\it m_{2}$ as a function of 
 $E_{th}^*$ for Au, La and Kr. The agreement between data and SMM 
is very good. The flatness of  $\it m_{2}$ beyond  $E_{th}^*$$\sim$ 8 MeV/nucleon observed for Kr 
is also seen in SMM. The size of the largest fragment from both data 
and SMM is shown
in Fig.\ref{amaxethsmm} for all the three systems. Very good agreement between the two is 
obtained over the entire range of $E_{th}^*$. 

\begin{figure}[ht]
\epsfxsize=8.5cm
\centerline{\epsfbox{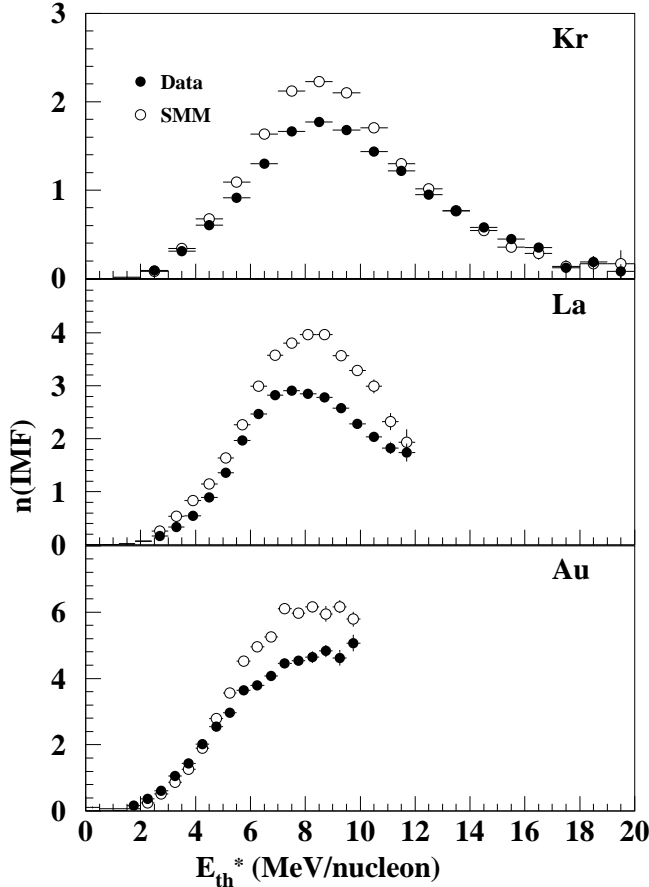}}
\caption{Average multiplicity of IMFs as a function of  $E_{th}^*$ 
for Kr, La, and, Au from data and SMM.
}
\label{imfethsmm}
\end{figure}

 
%
  The average value of the total number of IMFs
as a function of $E_{th}^*$ is shown in Fig.\ref{imfethsmm} for both data and SMM. 
Both show the same initial increase in IMF production and a peak at approximately the same energy. 
SMM follows the trend in the Kr data in the vaporization regime,
where the IMF multiplicity decreases to zero at the highest energy. 
 SMM overestimates the number of
IMF at the peak in all three cases. This difference is due to the fact that in SMM there is overproduction of Li and Be 
fragments at higher excitation energies. 

  The above comparison between data and SMM  was done with SMM 
calculations including the deexcitation of secondary fragments. SMM can also be stopped prior to secondary decay to obtain information about the 
SMM primary fragments. We refer to these results as
 $SMM_{hot}$ while those following 
 secondary decay are designated $SMM_{cold}$. As will be seen in the following sections $SMM_{hot}$
 can be used to evaluate the nature of the phase transition. 
A full description of $SMM_{hot}$ for the remnant system with A=160 and Z=64 has been given in ref.\cite{scharn2}.
\begin{figure}[ht]
\epsfxsize=8.5cm
\centerline{\epsfbox{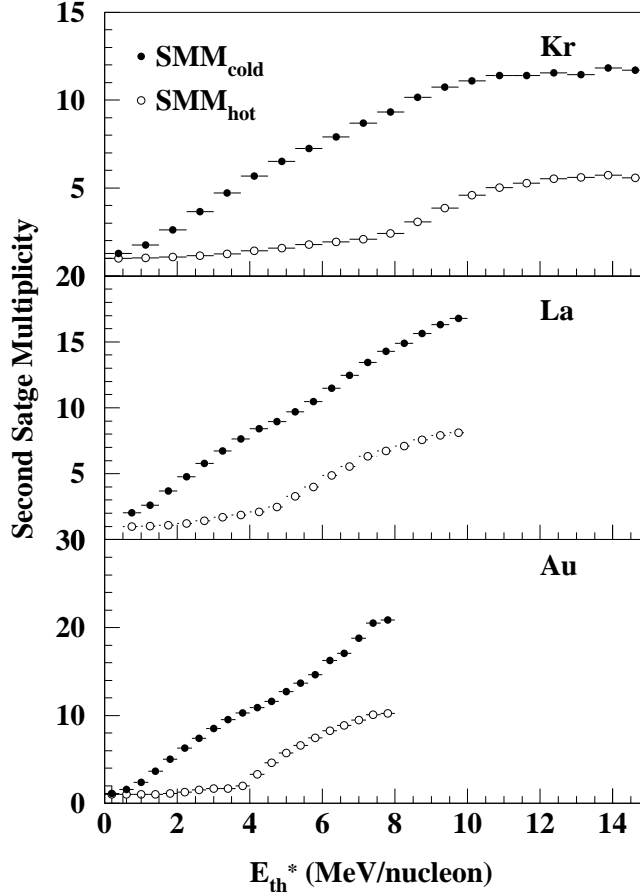}}
\caption{Average multiplicity from $SMM_{hot}$ and $SMM_{cold}$ as a function of  
$E_{th}^*$ 
for Kr, La, and, Au.
}
\label{smmhotcold}
\end{figure}
 The $ m_{2}$ distributions as a function of
 $E_{th}^*$  for both  $SMM_{cold}$ and $SMM_{hot}$ are shown in 
Fig.\ref{smmhotcold}. There 
is an  increase in multiplicity for  $SMM_{cold}$  in all three cases. For 
$SMM_{hot}$ $ m_{2}$ remains nearly unchanged up to a certain 
energy and then suddenly increases. The energy at which $ m_{2}$ starts
increasing corresponds to the MF threshold.
 The values of $E_{th}^*$ at which MF first occurs are $\sim$ 4, $\sim$ 5,
 and $\sim$ 8 MeV/nucleon for Au, La and Kr, respectively. The threshold energy is lowest for the heaviest
 system because of  
the large Coulomb energy in 
the heavier remnant, which facilitates the breakup of the nucleus at a 
lower energy.
 There is a very narrow window for Kr between the multifragmentation threshold and the energy at which vaporization starts. This reduces the 
probability of IMF formation, as is evident from Fig.\ref{imfeth}. 

\section{CRITICAL POINT DETERMINATION}
 
In our earlier publications we discussed the 1A GeV Au on carbon data in terms of the theory of critical phenomena and several critical exponents were 
determined \cite{gilkes94,elliott96,elliott98,elliott5,elliott6}. We
 used the percolation technique applied to small lattices to study 
 critical phenomena \cite{elliott94,elliott97,isis2}. The method of moments 
analysis was used by several groups 
\cite{kreutz93,bauer88,bao93,barz93,belkacem96,mastinu98,agostino99} to search 
for evidence of the liquid-gas phase transition in MF.
Recently, we used this method in the analysis of La and Kr
 data \cite{srivas00} using charged particle multiplicity as the order parameter. In this work $E_{th}^{*}$ is used as the order parameter. 
$E_{th}^{*}$
is a more fundamental parameter than multiplicity for comparing the three systems. 
Here, we first use the combination of moments to
find the signature of criticality in Au, La and Kr data.
For example, the reduced variance $\gamma_{2} = M_{2} M_{0}/M_{1}^2$, where 
$M_{1}$ and $M_{2}$ are the first and second moments of the 
mass distribution in an event and $M_{0}$ is the total multiplicity  including neutrons, is a useful quantity. The reduced variance  
$\gamma_{2}$ has  a peak value of 2
for a pure exponential distribution, $n_{A}\sim$ $e^{-\alpha A}$, regardless of
the value of $\alpha$, but $\gamma_{2}$  $>$ 2 for a power law
distribution, $n_{A}\sim$ $A^{-\tau}$, when $\tau > 2$ and the system is large
enough. Critical behavior requires that the peak value of
$\gamma_{2}$ be larger than 2\cite{campi1,campi2}.

\begin{figure}[ht]
\epsfxsize=8.5cm
\centerline{\epsfbox{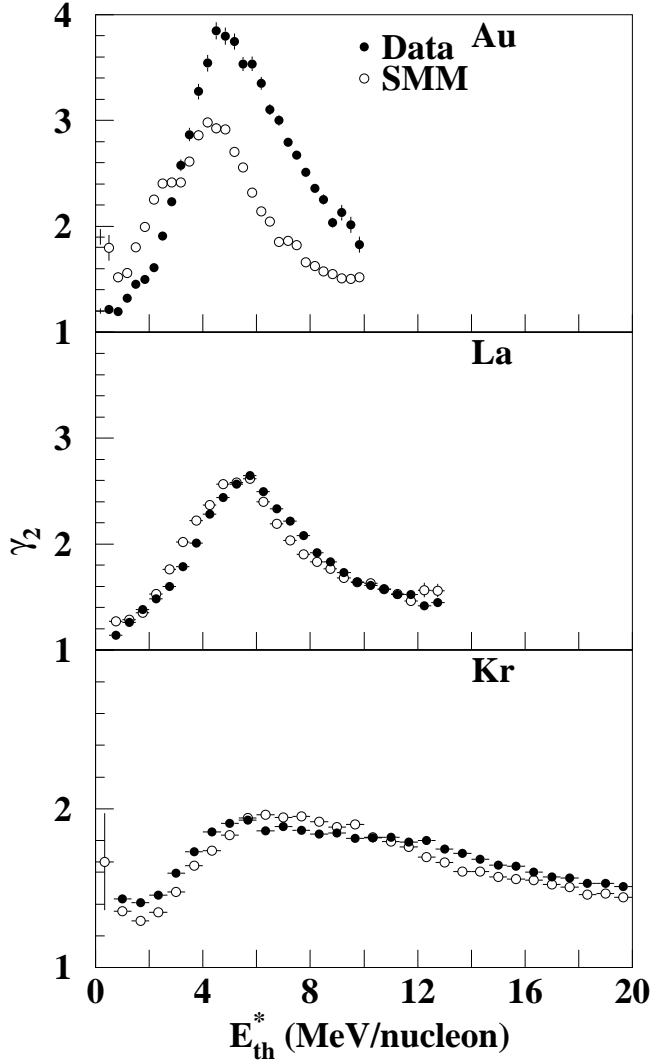}}   
\caption{$\gamma_{2}$ as a function of $E_{th}^*$ for all the 
three systems from data and $SMM_{cold}$. 
}
\label{gamma2einp}
\end{figure}
The event by event analysis for $\gamma_{2}$  
is shown in Fig.\ref{gamma2einp} for all the three systems. 
The position of the maximum $\gamma_{2}$ value defines the critical point,
 $\it i.e.$
the critical energy  $E_{c}^{*}$, where the fluctuations in fragment sizes are the largest. The 
peak in $\gamma_{2}$ is well defined for La and Au. The value for $E_{c}^{*}$
is in good agreement with our earlier values, where $m_{c}$ was used as the order parameter and $E_{c}^{*}$ was obtained from $E_{th}^*$ vs $\it m$ 
plot\cite{srivas00}. In case of Kr Fig.\ref{gamma2einp} shows that 
there is no well defined peak in $\gamma_{2}$ and the distribution is very broad. A well defined peak is obtained in $m_{c}$, but the value of 
$\gamma_{2}$ is always less than 2 \cite{srivas00}.
Fig.\ref{gamma2einp} also shows the  $\gamma_{2}$ calculation using $SMM_{cold}$. 
The fission contribution to  $\gamma_{2}$ has been removed both from the data analysis and SMM.
The $E_{c}^{*}$ values obtained for data and $SMM_{cold}$ are close to each other. There is a  difference in the height of the peak in $\gamma_{2}$ for Au between data and $SMM_{cold}$.  
It is also important to note that at the  peak $\gamma_{2} > 2$ both from $SMM_{cold}$ and data for Au and La and  $\gamma_{2} < 2$ for Kr.
This suggests that the exponent $\tau <2$ for Kr as there is no enhancement of $\gamma_{2}$ in the critical region . One expects an enhancement of the moments in the critical region with $\tau > 2$ as in 
most critical phenomena \cite{stauffer79,stauffer92}.

 In case of Au 
the $\gamma_{2}$ value remains above two for most of the excitation energy 
 range. The
$E_{th}^{*}$ width over which  $\gamma_{2}$ $>$ 2 is smaller for La and disappears for Kr. The decrease in  $\gamma_{2}$ with decrease in system size, as observed in 
Fig.\ref{gamma2einp}, is also seen in 3D percolation studies and the differences have been attributed to finite size effects \cite{campi92c,campi92}.
  The results for $\it E_{c}^{*}$ are given in Table I along with the values of
$\it m_{c}$ \cite{srivas00}.  
\begin{figure}[ht]
\epsfxsize=8.5cm
\centerline{\epsfbox{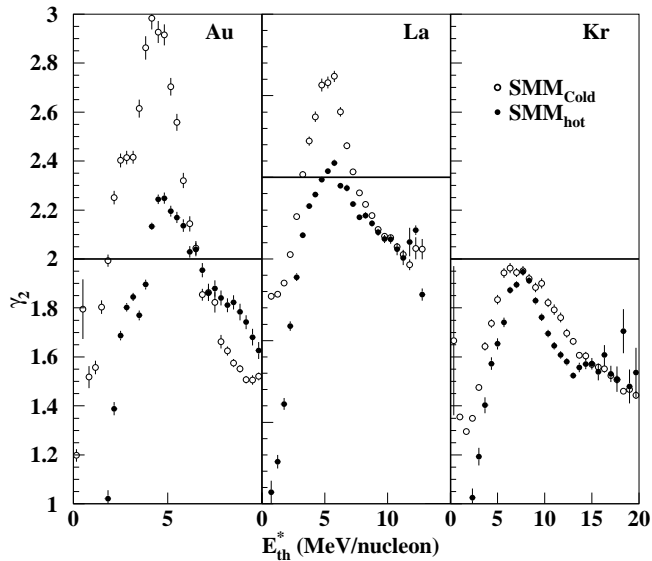}}   
\caption{$\gamma_{2}$ as a function of $E_{th}^{*}$ from $SMM_{hot}$ and 
$SMM_{cold}$ 
}
\label{gamma2hotcold}
\end{figure}
 It is of interest to see how the critical point determination from
$SMM_{hot}$ compares with that of $SMM_{cold}$. Since $SMM_{cold}$ can reproduce the various features of the EOS data, one can use $SMM_{hot}$ to 
understand the MF mechanism.  
 Fig.~\ref{gamma2hotcold} 
shows a plot of $\gamma_{2}$ 
as function of $E^*_{th}$ from $SMM_{cold}$ and $SMM_{hot}$. It is evident from
the plots for Au and La that the peak in $\gamma_{2}$ occurs at the same  
 $E^*_{th}$ for both $SMM_{cold}$ and $SMM_{hot}$. The distinct difference is in the height of the peak. For $SMM_{hot}$ the height of the $\gamma_{2}$ peak
is smaller as compared to  $SMM_{cold}$, but still above 2. This 
difference is mainly due
to the increase in $SMM_{cold}$ multiplicity as compared to  $SMM_{hot}$. 
In the case of Kr
the  $\gamma_{2}$ value is always less than 2 for both  $SMM_{cold}$ 
and $SMM_{hot}$ and the height of the $\gamma_{2}$ peak does not decrease from
  $SMM_{cold}$ to $SMM_{hot}$ as for Au and La. 
A detailed calculation of critical exponents from  $SMM_{hot}$ has been given in our earlier publication\cite{scharn2}. 
These results clearly demonstrate that $SMM_{hot}$ and 
 $SMM_{cold}$ behave in a similar way.
The most important conclusion is that the values of the exponents do not change in going from $SMM_{hot}$ to $SMM_{cold}$.
 Thus, $SMM_{hot}$ results are consistent with a critical phase transition for Au and La and not for Kr. 

\section{ CRITICAL EXPONENTS ANALYSIS}

\subsection{$\tau$ Exponent}

 In the previous section it was shown, based on the $\gamma_{2}$ analysis, that Au and La should show a power law  with exponent $\tau \ge 2$ for a continuous phase transition to be present. 
 $\tau$ can be obtained using the moments of the fragment mass distributions.

  Scaling theory \cite{stauffer79} relates the values of the critical exponents 
$\tau$ and $\sigma$ to the moments $M_{k}$ of the mass distribution through

\begin{equation}
M_{k} \propto |p-p_{c}|^{-(1+k-\tau)/\sigma}
\end{equation}
where p is the bond breaking probability and at the critical point p=$p_{c}$ in percolation.
The values of $\tau$ and $\sigma$ in the above equation are characteristic
for the specific class of phase transition. For a transition of the 3D percolation type $\tau=2.2$ and $\sigma=0.45$, while for a liquid-gas phase transition
a value of 
 $\tau=7/3$ and  $\sigma=2/3$. The 3D Ising values for  $\tau$ and $\sigma$ 
are 2.2 and 0.64 respectively. The $\tau$ value is essentially the same for 
different universality classes. The above equation can be solved 
to get the value of
$\tau$ if the second  $(M_{2})$ and third moments $(M_{3})$ of the 
fragment mass distributions are
known. A plot of $ln(M_{3})$ versus  $ln(M_{2})$ should give a straight 
line with a slope given by

\begin{equation}
 S= {\Delta ln(M_{3})\over \Delta ln(M_{2})} =  {\tau-4 \over \tau-3}
\end{equation}

\begin{figure}[ht]
\epsfxsize=8.5cm
\centerline{\epsfbox{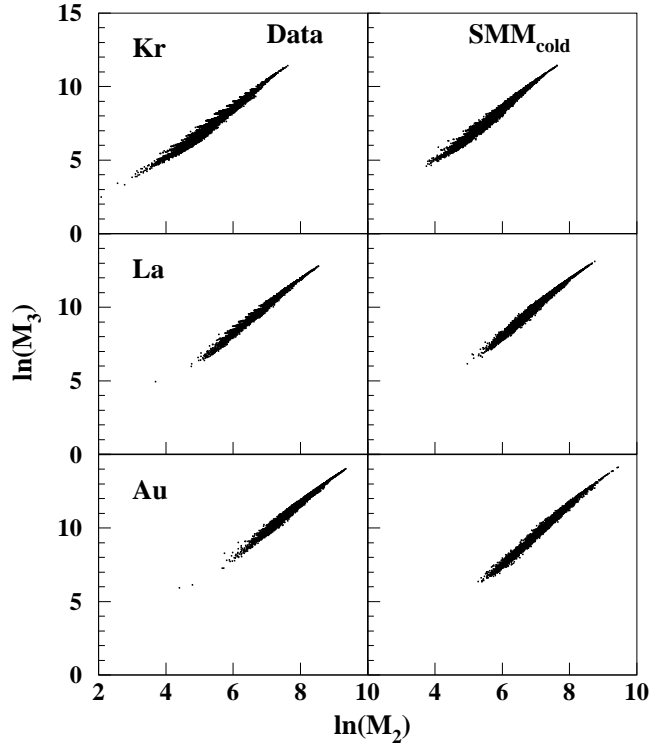}}   
\caption{$ln(M_{3})$ vs $ln(M_{2})$ for Au, La and Kr above the critical
 energy.}
\label{m3m2}
\end{figure}

Fig.\ref{m3m2} shows a scatter plot of $ln(M_{3})$ vs $ln(M_{2})$ for all the three systems from data above
$E_{c}^{*}$ \cite{elliott94}. Also shown in Fig.\ref{m3m2} are the results from $SMM_{cold}$. A linear fit to 
$ln(M_{3})$ vs $ln(M_{2})$ gives the value of $\tau$. The $\tau$ values are shown in Table I both from data and SMM.
A linear fit to the Kr data gives a value of $\tau =1.88\pm0.08$.
This value is below the minimum value $\tau \geq 2$ expected for a continuous 
phase transition. This result for Kr is consistent with Fig.\ref{gamma2einp}, 
which 
shows that $\gamma_{2}$ $<2$ and hence $\tau$ $< 2$\cite{campi1,campi2}. 
Similar results are also obtained for $SMM_{cold}$.

\begin{figure}[ht]
\epsfxsize=8.5cm
\centerline{\epsfbox{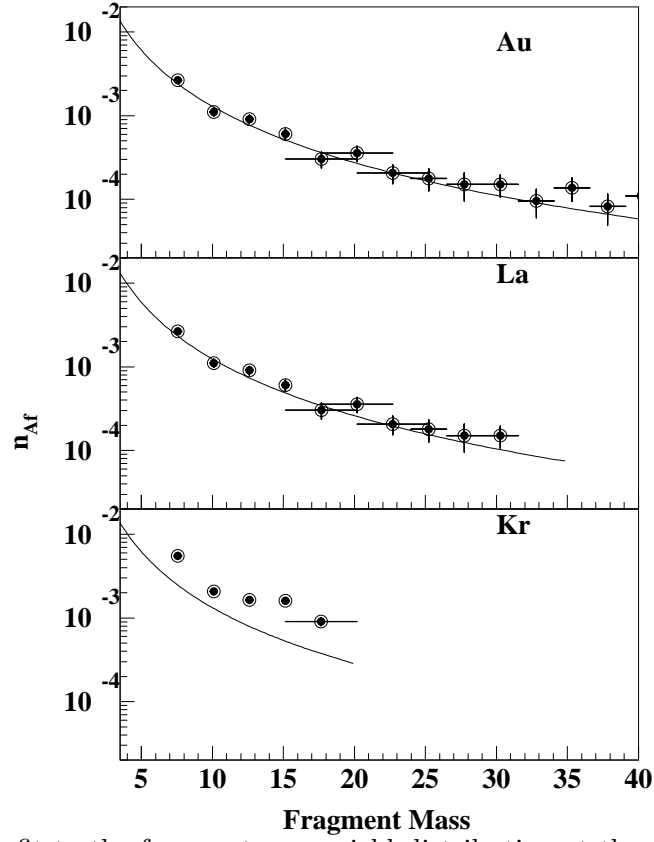}}   
\caption{Power law fit to the fragment mass yield distribution at the 
critical point. The solid line corresponds to 
$n_{Af}\sim q_{0}A^{-\tau}$, with $q_{0}$=0.2 and $\tau$=2.2[43]. 
}
\label{powerlaw}
\end{figure}

\begin{table}
\caption{Critical parameters  from data and $SMM_{cold}$}
\label{tab1}
\begin{tabular}{ccccccccccc}
Parameter&$Au_{data}$&$La_{data}$&$Kr_{data}$&$Au_{smm}$&$La_{smm}$&$Kr_{smm}$
&Per$^a$&LG$^b$\\
\tableline\\
$m_{c}$&28$\pm$3&24$\pm$3&18$\pm$2&26$\pm$3&23$\pm$3&17$\pm$2\\
$E_{c}^{*}$&4.5$\pm$0.5&5.5$\pm$0.6&6.5$\pm$1.0&4.3$\pm$0.5&5.3$\pm$0.6&6.2$\pm$1.0\\
$\tau$&2.16$\pm$0.08$$&2.10$\pm$0.06&1.88$\pm$0.08&2.11$\pm$0.05$$&2.05$\pm$0.05&1.81$\pm$0.06&2.20&2.21\\
$\beta$&0.32$\pm$0.02$$&0.34$\pm$0.02&0.53$\pm$0.05&0.35$\pm$0.03$$&0.37$\pm$0.03&0.57$\pm$0.06&0.44&0.328\\
$\beta/\gamma$&0.22$\pm$0.03$$&0.25$\pm$0.01$$&0.50$\pm$0.01&0.28$\pm$0.03$$&0.29$\pm$0.05$$&0.52$\pm$0.01\\
$\gamma$&1.4$\pm$$0.3^c$&-&-&1.02$\pm$$0.23^c$&-&-&1.76&1.24\\
$\gamma$&1.32$\pm$$0.15^d$$$&1.20$\pm$$0.08^d$\\
\end{tabular}
\begin{description}
\item a. Percolation,  b. Liquid-Gas,   c. ref.[47],  d. from $\beta$ and $\beta/\gamma$ ratio
\end{description}
\end{table}   
  A one-parameter power law search also provides an alternative method for the determination of $\tau$ \cite{elliott5}. At the critical point the cluster
distributions were described by $n_{Af}\sim q_{0}A^{-\tau}$ with
 $\tau \sim$ 2.2 and $q_{0} \sim$ 0.2, as seen in many universality
classes \cite{elliott5}.
 Fig.\ref{powerlaw} shows a plot of the fragment mass distribution at 
the critical energy for 
Au, La and Kr. 
The line corresponds to $n_{Af}= q_{0}A^{-\tau}$ with 
$\tau$=2.2 and $q_{0}$=0.2 \cite{elliott5}.  
It is clear from the plot that Kr
data do not follow the above power law with the critical values of $q_{0}$ and
$\tau$.

\subsection{$\beta$ exponent}
  
 Using the methods developed in percolation studies the value of the exponent 
$\beta$ can be obtained for the MF of 1A GeV Au+C \cite{gilkes94}. 
It is related to the
size of the largest cluster by the relation 

\begin{equation}
A_{max} \sim |\epsilon|^{\beta}
\end{equation}
where $\epsilon=p-p_{c}$ and $\epsilon>0$.
In the MF case $\it p$ and $\it p_{c}$ have been replaced by  
$E_{th}^{*}$ and $E_{c}^{*}$.
In the infinite lattice, the infinite cluster exists only on the liquid side of
$\it p_{c}$. In a finite lattice a largest cluster is present on both sides of
the critical point, but the above equation holds only on the liquid side. 
When $\epsilon<0$ , no infinite cluster exists and the size of the finite cluster is given by 
 \begin{equation}
A \sim |\epsilon|^{-(\beta+ \gamma)}
\end{equation}
where $\gamma$ is another critical exponent and related to the 
second moment, $M_{2}$
\begin{equation}
M_{2} \sim |\epsilon|^{-\gamma}
\end{equation}

\begin{figure}[ht]
\epsfxsize=8.5cm
\centerline{\epsfbox{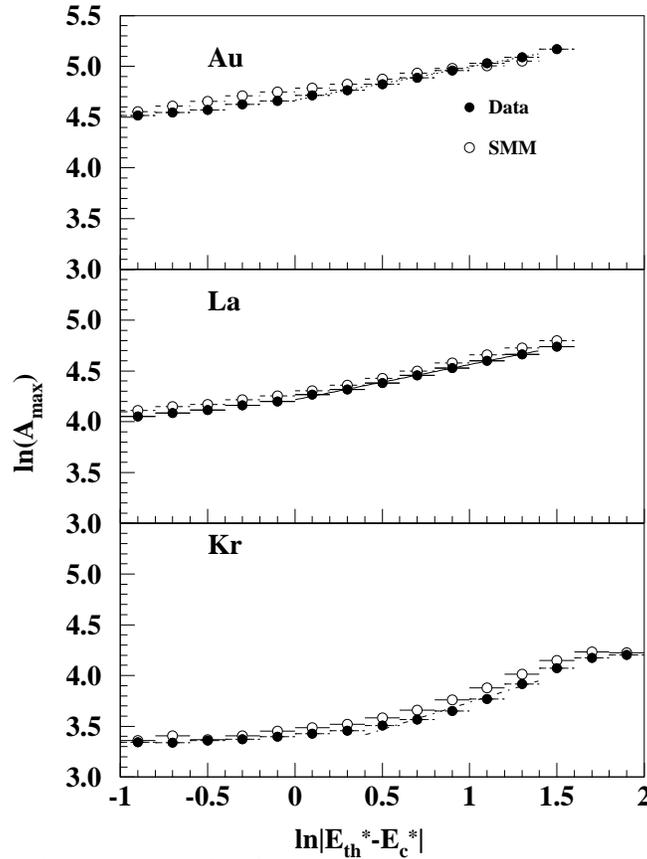}}   
\caption{$ln(A_{max})$ vs $ln|E_{th}^{*}-E_{c}^{*}|$ for Au, La and Kr below the  
critical energy for exponent $\beta$ determination. 
}
\label{beta}
\end{figure}

Fig.\ref{beta}
shows a plot of $ln(A_{max})$ vs $ E_{th}^{*}- E_{c}^{*}$ from Au, La and Kr. The values of $\beta$ obtained from the fit are given in Table I.  
The values of $\beta$ for Au and La are 0.32$\pm 0.02$ and 0.34$\pm 0.02$, 
respectively, and close to the value of 0.33 predicted for a liquid-gas 
phase transition.
 The value of $\beta$ for Au is in agreement with our earlier reported value\cite{gilkes94}. In case of La $\beta$ is close to the value obtained for Au, 
while the value of $\beta$= 0.53$\pm 0.05$ for Kr is much higher than that 
of Au and La. 

 It is interesting to note that the SMM calculation also gives the same results as obtained in data.
 Fig.\ref{beta} shows results from SMM as open circles for all the three systems. These comparisons with SMM are important, as this will
help us to probe the order of phase transition using SMM in the three experimentally studied systems. 

\begin{figure}[ht]
\epsfxsize=8.5cm
\centerline{\epsfbox{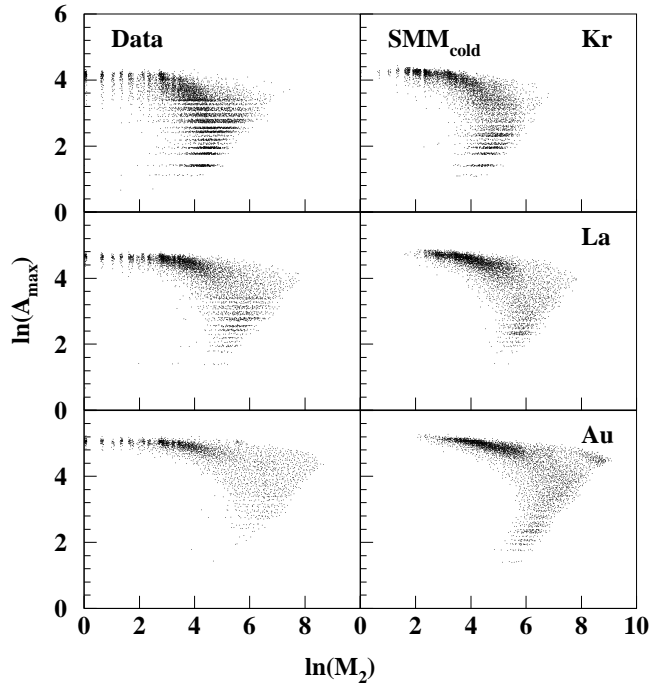}}   
\caption{Scatter plot of the $ln(A_{max})$ vs $ln(M_{2})$ from data and $SMM_{cold}$.
}
\label{amaxm2}
\end{figure}

 Campi\cite{campi1,campi2}  also suggested that the correlation between the 
size of biggest 
fragment, $A_{max}$, and the moments in each event can  measure
the critical behavior in nuclei. Fig.\ref{amaxm2} shows a plot of the logarithm of $A_{max}$ in each event versus the logarithm of the second moment $M_{2}$ for all the three systems. This plot is generally called a Campi scatter 
plot and has been 
successfully used in many studies \cite{jaqaman91,belkacem95}. The two branches corresponding to
the under-critical ( upper branch ) and over-critical (lower branch) events are clearly seen for Au and La. The $A_{max}$ -  $M_{2}$ correlation is quite broad for Kr and fills most of the
available phase space.
The lower and upper branches seems to overlap and are not well separated.
Studies on percolation lattices show similar behavior \cite{jaqaman91}.
However, it is not possible from such a plot to locate the critical region 
in a precise and unambiguous manner. But, if we know the critical 
point from some 
other method, then the $A_{max}$- $M_{2}$ correlation  can be used to calculate the ratio of critical exponents $\beta/\gamma$ from the slope of the 
upper branch. A similar behavior is obtained from SMM as shown in 
Fig.~\ref{amaxm2} for all the three systems. 
Table I shows the $\beta/\gamma$ values for
Au and La from the linear fit to the upper branch of Fig.\ref{amaxm2}. For the fitting purpose an average value of $ln(A_{max})$ was obtain for each value of 
$ln(M_{2})$.

 We have mentioned only a few exponents in this analysis and compared them with SMM. The aim is to show that Kr is different than Au and La. The two exponents $\tau$ and $\beta$ serve this purpose very well. Knowing $\beta$ and 
 $\beta/\gamma$, $\gamma$ can be obtained. Results are listed in Table I and 
 the value for Au agrees with the published value. 
 Table I also gives the  values of the critical exponents for the
 percolation and liquid-gas phase transition in 3D systems.  
No attempt has been made in this paper to independently determine other 
exponents like $\gamma$ and $\sigma$. 
 All the exponents for Au and their comparison with SMM have been reported 
in various publications\cite{gilkes94,elliott96,elliott98,scharn2}.

\section{ENERGY FLUCTUATIONS AND HEAT CAPACITY ANALYSIS}

  In a recent study experimental evidence of a
liquid-gas phase transition in MF was offered by analyzing the 
fluctuations in the total fragment energy \cite{chomaz99,agostino00}.
It has been shown that for a given total energy the average partial energy 
stored in a part of the microcanonical system is a good thermometer while 
the fluctuations associated with the partial energy can be used to determine 
the heat capacity, which is negative for a first order transition and 
positive for a second order transition. 
 \begin{figure}[ht]
\epsfxsize=8.5cm
\centerline{\epsfbox{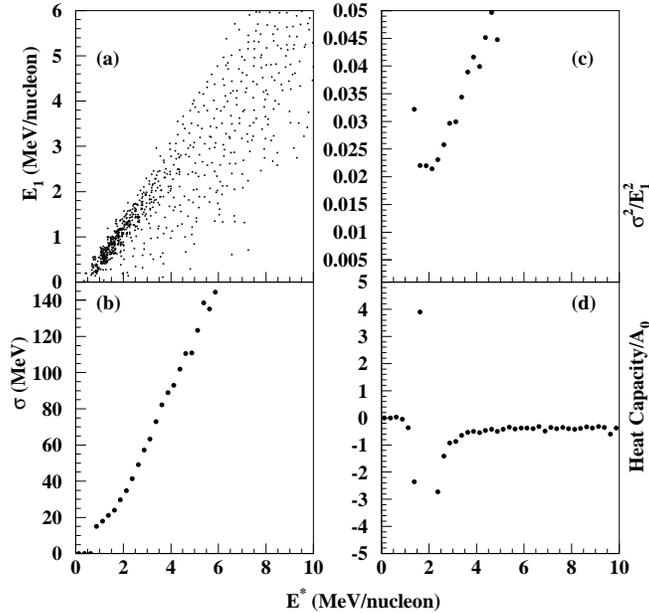}}   
\caption{With no selection on the remnant mass:
a). Partial energy $E_{1}$ as a function of $E^*$ 
b). Fluctuation in $E_{1}$ c). The variance ($\sigma/E_{1})^2$ and d). Heat capacity per nucleon for Au .  
}
\label{aucv}
\end{figure}
 A negative heat capacity was obtained in  the reaction Au+Au at 
35A MeV, providing the direct evidence of a first order liquid-gas phase 
transition \cite{agostino00}. We have analyzed our data for Au and Kr using energy fluctuations. 
From the  total excitation energy $E^*$ the Coulomb and expansion 
energy components 
were removed to obtain energy $E_{1}$ \cite{hauger98,hauger00}.
 \begin{figure}[ht]
\epsfxsize=8.5cm
\centerline{\epsfbox{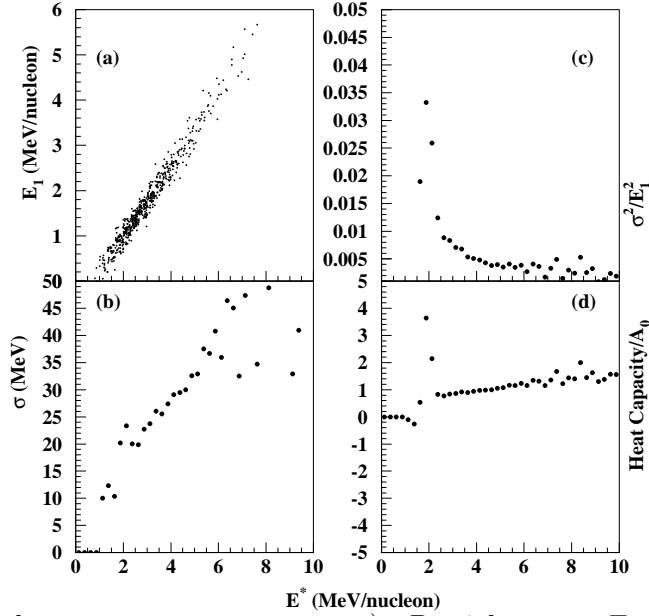}}   
\caption{With a tight cut on remnant mass: 
a). Partial energy $E_{1}$  as a function of $E^*$ 
b). Fluctuation in $E_{1}$ c). The variance $(\sigma/E_{1})^2$ and d). Heat capacity per nucleon for Au. 
}
\label{aucvcut}
\end{figure} 
The fluctuations in $E_{1}$ were studied as a function of  $E^*$.
Figs.~\ref{aucv}(a) and (b) show the spread in $E_{1}$ and its standard deviation 
$\sigma$ respectively, as a function of $E^*$. The
reduced variance ${(\sigma /E_{1})}^2$ is shown in Fig.~\ref{aucv}(c). The heat capacity $C_{t}$ is given by
\begin{equation} 
C_{t} = C_{1}^2/[{C_{1}-{(\sigma/T)}^2}] 
\end{equation}
where $ C_{1}$ is the canonical specific heat. It was argued that  a negative $C_{t}$ in the critical region is a signature of a 
first order phase transition\cite{chomaz99,agostino00}. 
In Fig.~\ref{aucv}(d) $C_{t}$ is shown as a function of $E^*$. There is a negative heat capacity at $E^*$ $\sim$ 2-3 MeV/nucleon.

However, a different result is obtained if there is a selection on the remnant mass. Figs.~\ref{aucv}(a) and (b) have no remnant mass 
cut and the
average mass is $170\pm 19$. This causes large fluctuations in $E_{1}$.
Figs.~\ref{aucvcut}(a)-(d) show similar plots as in Figs.~\ref{aucv}(a)-(d) with 
the remnant restricted to $161\pm3$. 
Two important results come out of these plots. First, 
 $\sigma$ is much smaller when the remnant mass cut is applied. Second, 
$C_{t}$ is always positive. Our analysis for Au data shows no negative heat capacity. Based on our 
statistical and thermodynamic analysis, we see indications for
a continuous phase transition in Au data\cite{gilkes94,elliott96,elliott98,scharn2}. The results for La are similar to those shown above for Au.

\begin{figure}[ht]
\epsfxsize=8.5cm
\centerline{\epsfbox{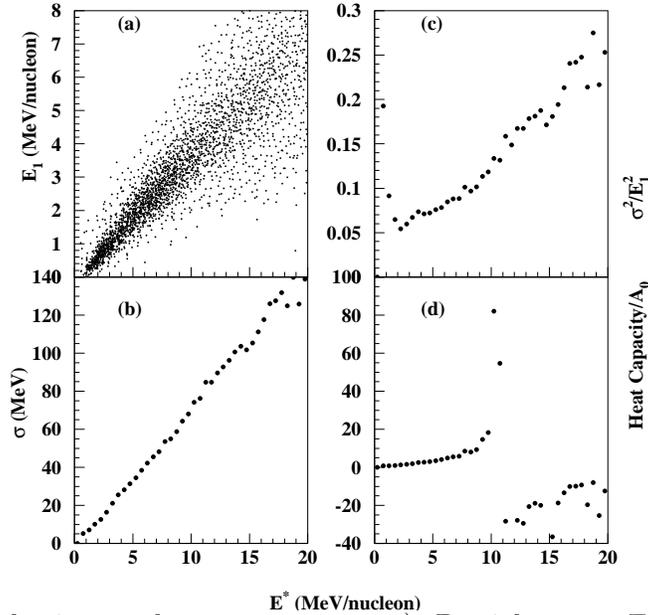}}   
\caption{With no selection on the remnant mass: 
a). Partial energy $E_{1}$ as a function of $E^*$
b). Fluctuation in $E_{1}$ c). The variance $(\sigma/E_{1})^2$ and d). Heat capacity per nucleon for Kr 
}
\label{krcv}
\end{figure}

\begin{figure}[ht]
\epsfxsize=8.5cm
\centerline{\epsfbox{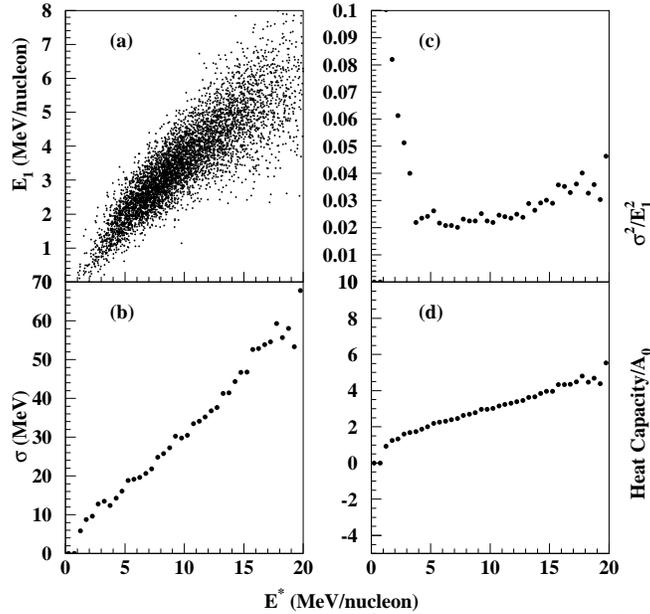}}   
\caption{With a tight cut on remnant mass: 
a). Partial energy $E_{1}$ as a function of 
$E^*$
b). Fluctuation in $E_{1}$ c).The variance $(\sigma/E_{1})^2$ and d). Heat capacity per nucleon for Kr 
}
\label{krcvcut}
\end{figure}

A similar analysis was performed on Kr data and results are shown in 
Figs.~\ref{krcv}(a)-(d) and Figs.~\ref{krcvcut}(a)-(d) 
for remnant masses $58\pm 14$ and $56\pm3$ respectively. We do not see the  signature of a first order transition even in Kr if the remnant mass is selected in a narrow bin. On the other hand a negative value of $C_{t}$ is obtained if there is no selection on the remnant mass.

Here, we emphasize again that the event-by-event complete 
 reconstruction  of remnant mass and excitation energy
is very important for the above analysis. This is possible only in high energy reverse kinematic 
asymmetric collisions \cite{hauger98}. In case of the EOS data, though there is a positive specific heat, 
 there are no large scale fluctuations in it to define the critical energy. Only  Au has a peak in $C_{t}$, which is around 2-3 MeV/A. There is no peak in  $C_{t}$ for La and Kr. Thus the analysis of EOS data for all three systems  based on heat capacity is questionable.   

\section{NATURE OF PHASE TRANSITION}

  In the previous sections we have shown that the MF of Kr is different than that of Au and La. Based on the experimental results it is not possible to 
decide the order of phase transition in Kr. The
 statistical analysis suggests a  continuous 
phase transition for Au and La. 

In our earlier publication \cite{scharn2} and in this paper as well we have shown that SMM can reproduce the various features of EOS data, including the
critical exponents. A single set of parameters of the model was used to describe the data for all the three systems, Au, La, and Kr. The nature of the phase transition in SMM was analyzed using $SMM_{hot}$, i.e. fragments formed before 
secondary decay \cite{scharn2}. The microcanonical temperature was obtained for the MF system and a caloric curve was constructed \cite{gross1,scharn2,huller94} for A=160, 130, 100, and A=70. A backbending in the caloric curve is an indication of a first order phase transition and leads to a negative 
specific heat. 
For A=160 there is a positive peak in the specific heat vs energy plot, consistent with our energy fluctuation analysis on our data, using a mass cut.  
 \begin{figure}[ht]
\epsfxsize=8.5cm
\centerline{\epsfbox{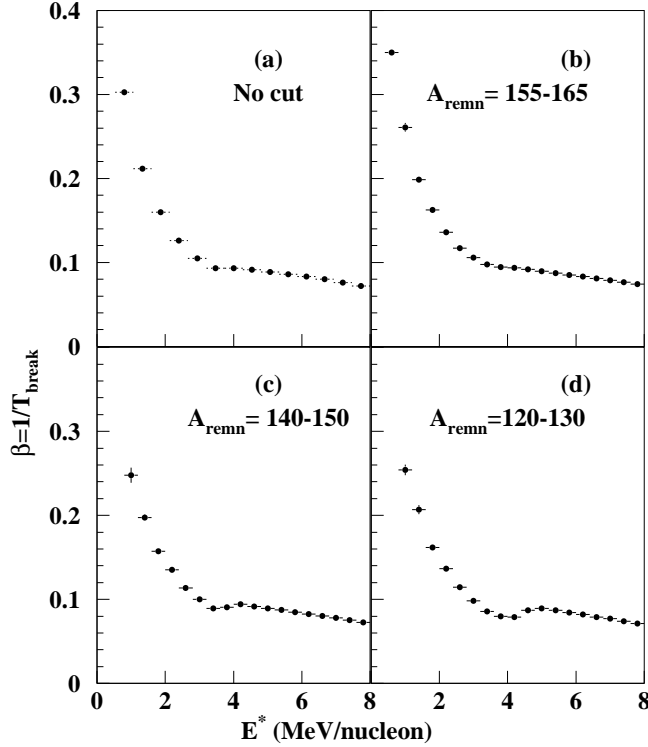}}   
\caption{ $\beta=1/T_{break}$ as a function of $E^*$ from $SMM_{hot}$ using experimental remnant as input to SMM. The graphs are shown for different mass selection
on the remnants from MF of Au.
}
\label{betaeth}
\end{figure}  

\begin{figure}[ht]
\epsfxsize=8.5cm
\centerline{\epsfbox{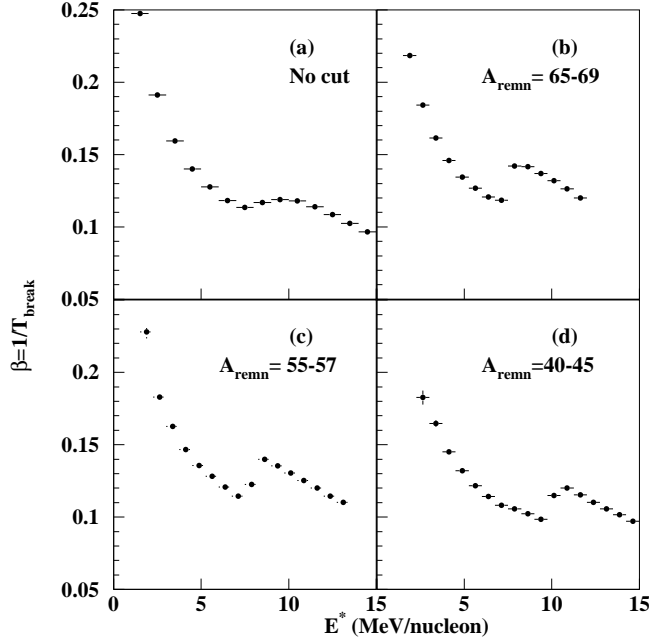}}   
\caption{ $\beta=1/T_{break}$ as a function of $E^*$ from $SMM_{hot}$ using experimental remnant as input to SMM. The graphs are shown for different mass selection
on the remnants from MF of Kr.
}
\label{betaeth2}
\end{figure}

  The back-bending in the caloric curve is sensitive to the remnant mass.
 We have used the remnant distributions from Au+C and Kr+C data to study the remnant mass effect on the caloric curve. Figs.\ref{betaeth}(a)-(d) show a plot 
of the reciprocal of the $SMM_{hot}$ breakup temperature
$\beta=1/T_{break}$, for Au. A distinct pattern is evident. A small back bending starts appearing at the very low mass cut.
 For heavier masses there is no back-bending. 
Since the remnant mass distribution
from Au+C interaction is dominated by  heavier remnants, the effect of
lighter remnants on back-bending is almost negligible. 
Fig.\ref{betaeth} suggests that, according to SMM, the MF transition in Au changes from first order to
second order depending on the remnant mass.  
For Kr, the back-bending is very sensitive to the particular mass cut. 
If there is no selection on mass cut the caloric curve is different than those with mass cuts as shown in 
Figs.~\ref{betaeth2}(a)-(d).
However, back-bending is always present. 
 Thus, based on the above reasoning we can say that in case of Kr there is
no continuous phase transition. Rather, analysis of the caloric curve argues 
for a first order phase transition in Kr. 

\begin{figure}[ht]
\epsfxsize=8.5cm
\centerline{\epsfbox{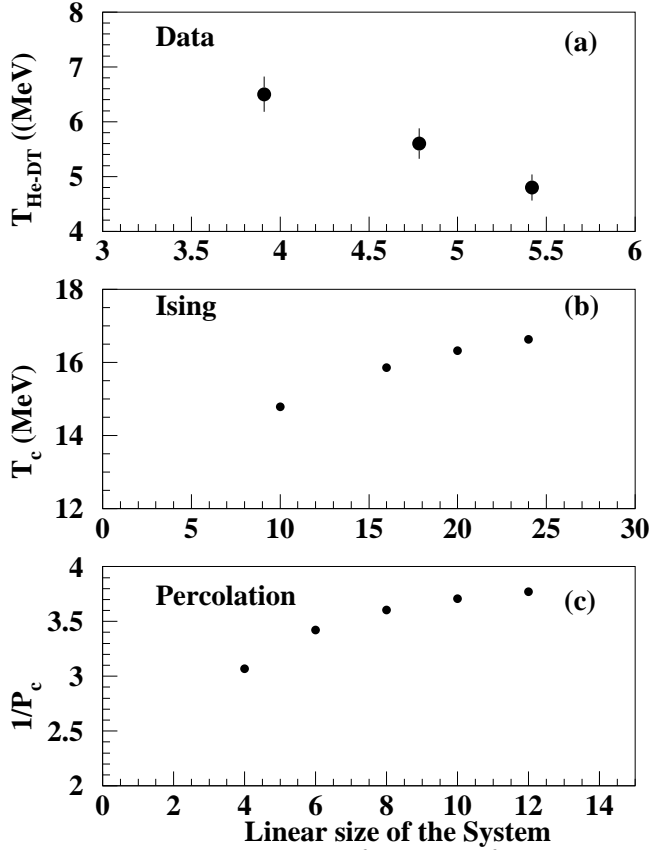}}   
\caption{a). $T_{He-DT}$ temperature at $E_{c}^{*}$ as a function of system size from data b). Critical temperature from Ising model calculation and c). Critical
bond formation probability plotted as $1/p_{c}$ for different lattice sizes.
}
\label{linear}
\end{figure}

 So far we have shown that the Kr data are different than the Au and La data.
 The temperature of the MF system provides further insight.
 One can obtain the freeze out temperature of the cold fragments in data 
using the ratio of light fragment isotopic yields \cite{hauger98,hauger00,albergo85}. In Fig.\ref{linear}(a) the isotope ratio temperature 
($T_{He-DT}$), as obtained from $^{2}H/^{3}H$  to  $^{3}He/^{4}He$  yield ratio at the 
critical point $E_{c}^{*}$,
is shown for Au, La and Kr systems as function of the linear size of the 
system (L= $A_{remn}^{1/3}$).
  $T_{He-DT}$ decreases with an increase in system
size. Recently, in an another study it has also been shown that critical 
temperatures and excitation energies decrease with increasing 
system size \cite{natowitz}.
This result can be attributed to the higher  Coulomb energy 
in Au as compared to Kr,
which in turn shifts the MF transition to lower temperature. The behavior of 
 $T_{He-DT}$ with system size is different than those observed either in percolation or Ising model studies. Fig.~\ref{linear}(b) shows a plot of the critical
temperature from 
a 3D Ising-type model with fixed density \cite{carmona98} for different lattice sizes. In the Ising calculation the trend is different than the one observed in Fig.~\ref{linear}(a). Thus, for neutral matter the critical temperature increases with an 
increase in system size. We can also compare the results from data and the Ising model calculation with  percolation studies. In percolation the critical 
probability $p_{c}$ in bond building percolation decreases with increase in the
system size\cite{elliott97,stauffer2}. Fig.~\ref{linear}(c) shows 1/ $p_{c}$ vs percolation lattice size, where  $p_{c}$ is inversely proportional to  $T_{c}$ \cite{stauffer2}. 
Thus, MF of Au, La and Kr is different than 3D Ising and percolation models.
As mentioned earlier, the finite size affects only non-thermodynamical quantities e.g. $\gamma_{2}$, $A_{max}$, etc. and not energy or temperature. 

\begin{figure}[ht]
\epsfxsize=8.5cm
\centerline{\epsfbox{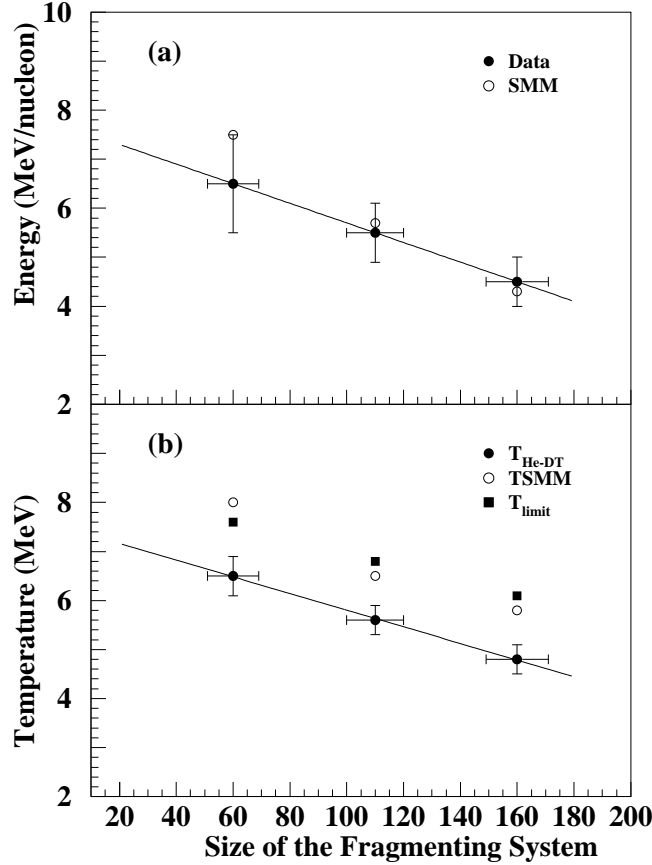}}
\caption{
a). Energy (MeV/nucleon) at critical point. b). $T_{He-DT}$,
$T_{SMM}$ and $T_{limit}$ as a function of the  system size.
}
\label{caloric}
\end{figure}

 The critical energy
is shown in Fig.\ref{caloric}(a) as a function of system size. The decrease in $E_{c}^{*}$
with increase in system size is evident from the figure. SMM also predicts the critical energy of the MF transition and this energy
is shown in Fig.\ref{caloric}(a) along with 
the data. This energy is
obtained at the hot fragment stage from the peak in $\gamma_{2}$ as shown 
in Fig.\ref{gamma2hotcold}.
Both data and SMM $E_{c}^{*}$ are in good agreement. The breakup temperature $T_{SMM}$, 
as obtained from $SMM_{hot}$, along 
with the $T_{He-DT}$ from data is shown in Fig.\ref{caloric}(b). There is a decrease in both temperatures with increase in system size. It is apparent that $T_{He-DT}$
is about 1 MeV lower than the SMM temperature. This difference is due to the fact that $T_{He-DT}$ is measured after secondary decay has taken place, while 
$T_{SMM}$ corresponds to the breakup configuration. Note
 that $T_{He-DT}$ tracks $T_{SMM}$ with system size at the 
critical point. SMM indicates that the decrease in both $T_{SMM}$ and 
$E_{c}^{*}$ with increasing system size is due to the increase of Coulomb 
energy. This result suggests that Coulomb energy plays an important role in
the MF of nuclei. It is interesting to note that the critical energy obtained
both  from SMM and data corresponds to the MF threshold as shown in Fig.\ref{smmhotcold}. The significance of this result remains to be determined. 

 The effects of finite size and Coulomb force on the MF have been studied by several workers and it is found that the decrease in critical temperature 
with increase in system size is primarily due to the 
Coulomb energy\cite{jaqaman84,bonch85,levit85}. The result from one such calculation is shown Fig.\ref{caloric}(b) as $T_{limit}$\cite{levit85}.   

The microcanonical Metropolis Monte Carlo (MMMC) \cite{gross1,gross2} calculations have 
emphasized that  MF is controlled by the competition between long range
Coulomb forces and finite size effects ( especially surface energy).
Finite size effects in models with only short range forces predict an 
increase in the critical temperature as the system size increases, as is evident from
percolation\cite{stauffer2} and Ising model studies\cite{carmona98}
 (see Fig.\ref{linear}). 
Since the experimental temperature 
exhibits the opposite dependence on system size, it is apparent that
Coulomb effects are more important than finite size effects.
For finite $ \it neutral$ matter the critical temperature ($T_{c}$)
 is expected to be $\sim$ 15-20 MeV \cite{jaqaman84,lattimer85}.
 The observed $T_{c}$ for A=160 is $\sim$ 6 MeV. 
Compared to finite uncharged nuclei, the presence of Coulomb energy plays
a role in lowering the excitation energy needed to reach the
regime where critical signatures are observed. In the smaller
Kr system there is less Coulomb energy in the initial remnant state.
Achieving multifragmentation in this system requires greater excitation
energy/nucleon compared to Au and La (as shown in Fig.~\ref{caloric}(a)) and as a result, the
dynamics of the ensuing disassembly may not take this system near its
critical regime. 
\begin{figure}[ht]
\epsfxsize=8.5cm
\centerline{\epsfbox{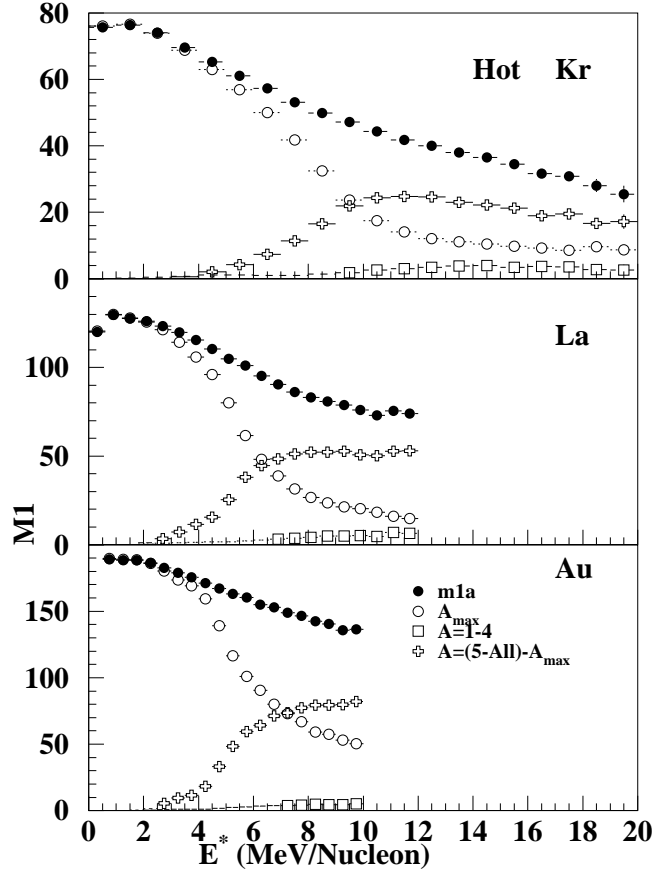}}
\caption{
M1 distribution for $SMM_{hot}$ for Au, La and Kr.
}
\label{m1hot}
\end{figure}

\begin{figure}[ht]
\epsfxsize=8.5cm
\centerline{\epsfbox{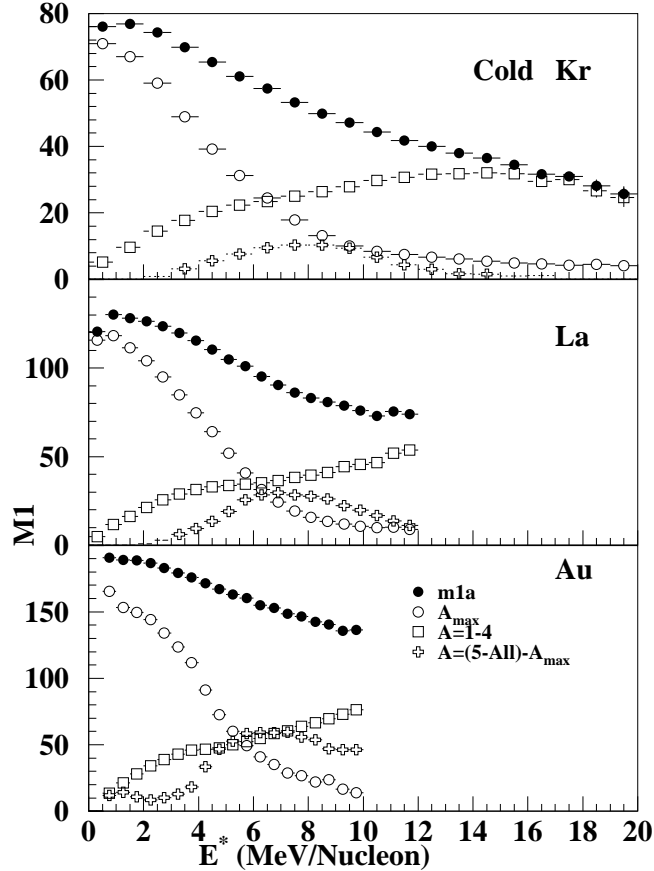}}
\caption{
M1 distribution for $SMM_{cold}$ for Au, La and Kr.
}
\label{m1cold}
\end{figure}

 It has also been suggested on the basis of a caloric curve that  MF of Au is a first order phase transition \cite{poch95}. 
This observation was based on the the fact that the temperature remains constant as energy is
increased between 3 and 10 MeV/nucleon. The temperature again starts increasing beyond 10 MeV/nucleon as a gas phase consisting of a mixture of nucleons and few light particles is created. According to SMM there are few nucleons in the
hot system.  
 Fig.\ref{m1hot} shows a plot of first moments of the fragment yield distribution
 as a function of  $ E^{*}_{th} $ for Au, La and Kr. The figure shows that even for  $ E^{*}_{th}$ $\geq $
10 MeV/nucleon very few particles with $A \le 4$ are produced. Most of the remnant mass is in fragments with $ A > 4 $. This argues against the coexistence of the constant density liquid and gas phases in SMM. The cooling of the hot fragments produces a large number of final state nucleons and light particles 
as shown in Fig.\ref{m1cold}. The IMFs survive the cooling process at least for Au and La and identify the MF transition. In case of Kr the excitation energies of IMF from $SMM_{hot}$ are very high and very few are seen in the cold stage. Thus the MF signal could be washed out. 
  
\section{SUMMARY AND CONCLUSIONS}

 In the present work we presented and analyzed the data from the MF of 
1A GeV Au, La and Kr on carbon. The mass, charge and excitation energy of the  remnant were determined in each event. The thermal excitation energy was
obtained after the expansion energy was determined on the basis of energy 
balance.
 The multiplicity distribution from MF of Kr shows a saturation beyond $E_{th}^{*}$ of 8 MeV/nucleon, indicating that the vaporization process has started. 

 A comparison of the data with the variable volume version of SMM, 
using the same values of parameters of the model for Au, La and Kr show a very good agreement with various distributions
e. g. charged particle multiplicity, size of biggest fragment, number of IMFs, 
etc. The power law behavior seen in data for Au and La with $\tau > 2$ is also observed in SMM when cold fragments are analyzed. The hot fragment analysis from $SMM_{hot}$ for $\gamma_{2}$ gives a value of $E_{c}^{*}$ which is in 
agreement with the value of $E_{c}^{*}$ obtained from $SMM_{cold}$. 

 The Au and La data give indications of a continuous phase transition. 
A first order phase transition has been predicted for Kr using SMM. 
However, the data cannot be used to distinguish between the two because
of finite size effects.
 
The SMM calculation offers
the best way to determine the the nature of phase transition. A back bending in
temperature vs energy plot is a sign of negative specific heat and hence a 
first order phase transition. Such a result is found for Kr but not for Au, except for the lightest remnants, the MF contribution of which is negligible.

 The temperature obtained from the isotope ratio analysis decreases with an
increase in system size at the critical point. The break-up temperature obtained from SMM also follows the same trend, but higher by $\sim$  1 MeV/nucleon.
The decrease in temperature with increase in system size is an important result as it is in the opposite direction from what is observed in either in 3D percolation or Ising model studies for finite size neutral matter. This shows that if finite size were the only effect in MF then the correlation between 
temperature and system size would have been the same in data as in percolation and Ising model studies. Thus, the long range Coulomb force is shown to be
the dominant factor in MF. This conclusion is also supported by Hartree-Fock calculations.   

 In conclusion, this is the first work in which the nature of the phase transition in MF has been 
explored using three systems of different size. The experimental results in conjunction with SMM provide the order of the phase transition in Au, La and Kr. The values of critical exponents $\tau$, $\beta$ and $\gamma$, which are close to the values for liquid-gas system, along with nearly zero latent  
heat suggest 
a continuous phase transition in Au and La. The back bending in the caloric curve for Kr suggests the presence of latent heat, which is consistent with a first order phase transition. We emphasize again here the important role played by the Coulomb energy. The Coulomb expansion energy reduces or eliminates the latent heat and changes the nature of the phase transition.

This work was supported by the U. S. Department of Energy.

\end{document}